\documentclass[arxiv]{revtex4}
\usepackage{amsmath,units,xspace,amssymb,tcolorbox}
\begin{document}

\title{Assembly and Phase Transitions within Colloidal Crystals}
\author{Bo Li,$^{1}$ Di Zhou,$^{1,2}$ and Yilong Han$^{1,*}$}
\affiliation{1, Department of Physics, Hong Kong University of Science and Technology, Clear Water Bay, Hong Kong, China. \\2, Multi-disciplinary Materials Research Center, Frontier Institute of Science and Technology, Xi'an Jiaotong University, Xi'an 710049, China.}
\date{\today}

\begin{abstract} 

 Micrometre-sized colloidal particles can be viewed as large atoms with tailorable size, shape and interactions. These building blocks can assemble into extremely rich structures and phases, in which the thermal motions of particles can be directly imaged and tracked using optical microscopy. Hence, colloidal particles are excellent model systems for studying phase transitions, especially for poorly understood kinetic and non-equilibrium microscale processes. Advances in colloid fabrication, assembly and computer simulations have opened up numerous possibilities for such research. In this Review, we describe recent progress in the study of colloidal crystals composed of tunable isotropic spheres, anisotropic particles and active particles. We focus on advances in crystallization, melting and solid–solid transitions, and highlight challenges and future perspectives in phase-transition studies within colloidal crystals.

\end{abstract}

\maketitle

Micrometre-sized colloidal particles can assemble into numerous structures and phases (FIG.1), and are excellent model systems for the study of phase transitions \cite{1990CRC_Ackerson, 1994PR_Lowen, 1996ARPC_Grier, 2002N_Lekkerkerker, 2006S_Frenkel,2002S_Frenkel} because their thermal motions can be directly visualized by optical video microscopy \cite{1973JCIF_Hachisu} and tracked by image processing \cite{1996JCIS_Crocker}. By contrast, the small spatial and timescales of atomic motions make the microscopic kinetics during phase transitions difficult to resolve. 

Phase transitions are ubiquitous in nature and industry, and have an important role in materials science, statistical physics, cosmology, biophysics, chemistry and earth science. They depend strongly on dimensionality, surface properties, defects, heating and cooling rates, and external fields. Phase transitions can be classified into first-, second- and infinite-order (in which the free energy is an infinitely differentiable exponential function such as the Kosterlitz–Thouless (KT) transition). The equilibrium behaviour of continuous phase transitions, including second- and infinite-order transitions, can be well described theoretically; however, the fundamental theory to describe first-order phase transitions is lacking. The kinetics of first-order and continuous phase transitions are  difficult to predict \cite{1987RPP_Binder}.

\begin{figure}[!h]
\centering
\includegraphics[width=0.6\columnwidth]{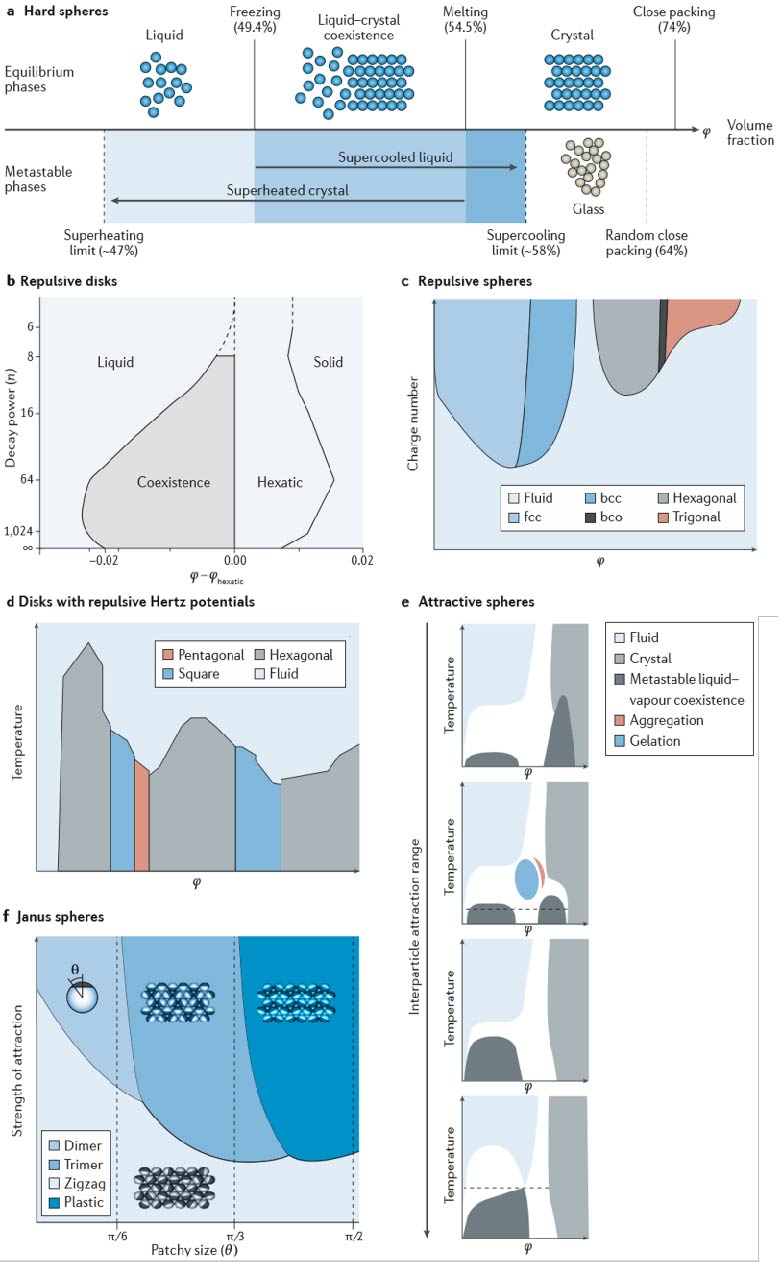}
\caption{\textbf{Phase diagrams of monodispersed colloids.} (a) Hard spheres. (b) Repulsive disks with pair potential $U(r)\sim r^{-n}$. The liquid–hexatic coexistence is a signature of first-order phase transitions. (c) Spheres with long-range repulsions for ionic colloidal crystals. (d) Spheres with repulsive Hertz potentials show an infinite number of cycles of 2D crystal phases. (e) Spheres with different attraction ranges. (f) Janus spheres with different patch sizes $\theta$ and interaction strengths. $\phi$, volume fraction; $\phi_{hexatic}$, volume fraction at the onset of the hexatic phase; bcc, body-centred cubic; bco, body-centred orthorhombic; fcc, face-centred cubic. Panel b adapted with permission from REF.154, American Physical Society. Panel c adapted with permission from REF.19, American Physical Society. Panel d adapted with permission from REF.21, Royal Society of Chemistry. Panel e adapted with permission from REF.5, American Association for the Advancement of Science. Panel f adapted with permission from REF.92, Royal Society of Chemistry.}
\label{fig:1}
\end{figure}

Colloidal particles can be viewed as large atoms with tailorable size, shape and interactions \cite{2002S_Frenkel}. Although colloids and atoms differ in some aspects (BOX 1), their phase transitions share many similarities; for example, they both follow classical nucleation theory (CNT) at weak supersaturations \cite{2012S_Han} and deviate from CNT at strong supersaturations \cite{2015NC_Han}. Hence, colloids can provide general information about phase transitions \cite{2002N_Lekkerkerker, 2006S_Frenkel,2002S_Frenkel}. Besides being important as micrometre-scale analogues of atoms, colloids are interesting in their own right. In the past two decades, colloidal model systems have provided vital insights into the microscopic kinetics of melting, crystallization, glass transitions and solid– solid transitions. Recent breakthroughs in colloidal fabrication, assembly and computer simulations open new opportunities for phase-transition studies. 

\begin{tcolorbox}

Comparison between colloidal particles and atoms: \\
Both colloidal particles and atoms experience strong thermal motions, which enable them to form thermally equilibrated phases. The phase behaviour of colloids with weak interactions, such as hard spheres, is mainly determined by the volume fraction, $\phi$, which is analogous to the effective inverse temperature, $1/T$, in atomic systems. Both a colloidal particle and an atom have $\sim k_BT$ thermal energy according to the equipartition theorem, but their diameters are very different (by a factor of $10^3$  or more). Consequently, the energy density and elastic moduli of 3D colloidal crystals are $10^9$ times smaller than those of atomic crystals (that is, they are `soft' materials). Unlike atoms, which exhibit ballistic motion in vacuum, colloidal particles undergo over-damped Brownian diffusion without a well-defined velocity. Colloidal particles dispersed in liquid have complex long-range hydrodynamic interactions in 3D, which influence the kinetics but not the equilibrium behaviour \cite{2010PRL_Tanaka, 2014PNAS_Crocker,2014EPL_Schilling, 2014SM_Arnold}. Note that a liquid solution is needed; otherwise microparticles form powders without Brownian motion. Quantum effects are usually negligible in colloids. Finally, colloidal particles can be made active with self-propelled motions, whereas atoms are passive.

\end{tcolorbox} 

In the first part of the Review, we describe conventional colloidal crystals composed of isotropic spheres that mimic simple atomic crystals, non-conventional colloidal crystals composed of anisotropic particles that resemble atoms with directional interactions or molecules, and crystals of active particles that have no counterpart in atomic systems. In the second part of the Review, we discuss recent studies of crystallization, melting, solid–solid transitions and other phase transitions in crystals composed of isotropic particles. Transitions within crystals composed of anisotropic or active particles are also briefly discussed. Owing to the limited space, we have highlighted only a limited number of examples of this rapidly expanding research field and focus on micrometre-sized particles.

\section{Colloidal crystals}

\subsection{Crystals composed of isotropic spheres}

Different particle shapes and interactions give rise to different packing entropy, $S$, and interaction energy, $U$, in minimizing the free energy, $F = U - TS$, where $T$ is temperature, thus resulting in different phases at equilibrium.
	
As the simplest colloid, monodisperse (that is, uniform size and properties) hard spheres can pack into face-centred cubic (fcc), hexagonal close packed (hcp) and random hexagonal close packed (rhcp) crystals and glasses in 3D \cite{1986N_Megen,1957JCP_Alder,2015S_Manoharan2}. Face-centred cubic packing makes most efficient use of the 3D space, and thus spheres have more effective room to move; that is, they have greater free-volume entropy, which can compensate for the extra configurational entropy of random packing. Consequently, the fcc crystal is the equilibrium phase with the maximum $S$ (that is, minimum $F$ \cite{1997N_Frenkel}) because $U = 0$ for hard spheres (FIG. 1a). In practice, however, hard spheres can be trapped in a supercooled liquid or a glassy state because the 3D locally favoured packing is a tetrahedron of four spheres that cannot tile the 3D space. In general, such frustration between locally and globally favoured packing can dynamically arrest the system into a disordered structure \cite{2002N_Lekkerkerker}. By contrast, hard or soft repulsive disks can form crystal and hexatic phases (FIG. 1b), but can not form a glassy phase in 2D even under fast quenching because the local optimal packing is an equilateral triangle of three disks which, can fully tile the 2D space. Crystals in lower dimensions have more long-wavelength fluctuations \cite{1988RMP_Katherine}, fewer neighbours to bind each particle on its lattice site and fewer interparticle `springs' connected in parallel rather than in series. Hence, crystals are too soft to exist in 1D unless there are long-range interactions; 2D is the critical dimensionality in which crystals are very soft with only quasi-long-range translational order and long-range orientational order. Between 2D crystals and liquids, there may exist a hexatic phase characterized by short-range translational order and quasi-long-range orientational order \cite{1988RMP_Katherine}. Hard spheres under quasi-2D confinement can form triangular, square and buckled thin-film crystals \cite{2006JPCM_Dijkstra, 1996PRL_Lowen}. 

Soft particles — that is, particles with long-range repulsions — can form fcc, body-centred cubic (bcc) or other lattices \cite{2004PRL_Lowen2, 2003N_Blaaderen} (FIG. 1c). Interestingly, continuously compressing soft Hertz-potential spheres can induce not only quasicrystals with five-fold symmetry but also infinite cycles of hexagonal and square crystals in 2D \cite{2011SM_Cacciuto} (FIG. 1d), and infinite cycles of fcc, bcc, hexagonal, simple cubic and body-centred tetragonal (bct) crystals in 3D \cite{2009JCP_Frenkel}. Pressure can further deform particle shape and induce new forms of crystals \cite{2010PRL_Miller}. Experimentally, particles can be softened by grafting long polymers, reducing the crosslinker density in microgels \cite{2009SM_Branka} or reducing the ionic strength to increase the Debye length of the screened Coulomb interaction for charged particles \cite{2004PRL_Lowen2, 2003N_Blaaderen, 2005N_Blaaderen}. Complex repulsions involving two or more length scales can result in various superlattices \cite{2013PRL_Buzza} and quasicrystals \cite{2014PRL_Lifshitz,2015NM_Glotzer,2014N_Ziherl}. 

Attractive colloidal particles can better mimic atoms. Attractive van der Waals forces universally exist among colloidal particles. The van der Waals force is strong over short distances ($<100$ nm) for microparticles and induces aggregation. To avoid such coagulation, van der Waals attractions are usually weakened below kBT by grafting polymers or surfactants to form a low-density surface (steric stabilization) or screened by electrostatic repulsions for charged particles (charge stabilization) \cite{2007SM_Yethiraj}. Beyond the van der Waals attraction range, typical interparticle attractions include depletion attraction induced by adding small particles \cite{2006S_Dinsmore, 2014NM_Chaikin, 2015PNAS_Irvine,2009PRL_Dinsmore}, and Casimir-like or wetting attraction induced by the fluctuation of either the two-component liquid near its critical point or percolation clusters of small particles \cite{2008N_Bechinger,2013NC_Schall,2015PRL_Dijkstra,2015NC_Sciortino}. Both the strength and range of attraction affect the phase behaviour (FIG. 1e). For example, when the range of attraction is much smaller than the repulsive core, particles form a fluid without distinguishable liquid and gas phases in colloida \cite{2006S_Frenkel} and molecular systems such as buckyballs \cite{1993N_Frenkel}. 
	
Unlike repulsive or long-range attractive colloidal spheres, which can easily form large crystals, shortrange attractive colloids often form disordered aggregates or gels \cite{2002N_Lekkerkerker} (FIG. 1e), indicating a rugged free-energy landscape. This can be explained in terms of the competition between locally and globally favoured structures \cite{2008NM_Tanaka}. When a liquid-like particle crystallizes, its neighbours must move to make way for new neighbours. Neighbours moving radially under short-range attraction require high energy, and tangential movement is often forbidden because of the constraints from neighbours' neighbours. Hence, the system tends to be deeply trapped in a disordered state.

Unlike atoms, colloidal particles cannot be perfectly identical in size or other properties. This polydispersity is a generic feature of colloids that can hinder crystallization and even induce new types of crystals \cite{2003PRL_Sollich,2015SP_Xu,2010S_Islam,2013PRE_Chen}. Hard spheres with $~6\%$ polydispersity can fractionate into crystallites with different mean diameters and undergo a re-entrant melting transition \cite{2003PRL_Sollich}. Hertz spheres with a polydispersity $<6\%$ can form normal crystals at high volume fractions, $\phi$, and distinct disordered crystals at low $\phi$ \cite{2015SP_Xu}. Uniformsized spheres with different softnesses can pack into another type of disordered crystal with perfect lattices and glassy behaviour \cite{2010S_Islam,2013PRE_Chen}. 

\begin{figure}[!h]
\centering
\includegraphics[width=0.8\columnwidth]{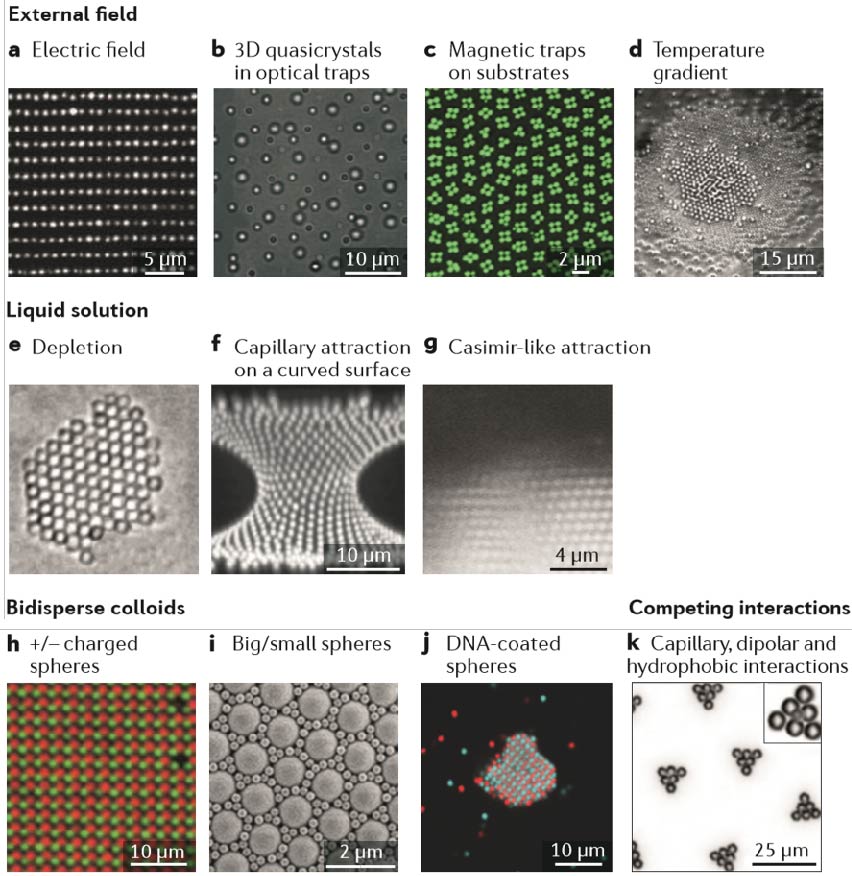}
\caption{\textbf{Strategies used to assemble isotropic particles into crystals.} (a-d) Application of external fields. (e-g) Adjustment of the liquid solution. (h-j) Use of bidisperse spheres. (k) Competing interactions. Image a adapted with permission from REF.43, American Physical Society. Image b adapted with permission from REF.45, Optical Society of America. Image c from REF.50, Nature Publishing Group. Image d adapted with permission from REF.46, Optical Society of America. Image e from REF.32, Proceedings of the National Academy of Sciences. Image f from REF.143, Nature Publishing Group. Image g from REF.34, Nature Publishing Group. Image h from REF.25, Nature Publishing Group. Image i adapted with permission from REF.47, Wiley–VCH. Image j adapted with permission from REF.95, American Association for the Advancement of Science. Image k adapted with permission from REF.26, American Physical Society.}
\label{fig:2}
\end{figure}

Compared with nanoparticles, microparticles do not self-assemble into high-quality crystals as easily, and defects are common. Hence, it is necessary to direct assembly by using an external force \cite{2004PRL_Blaaderen,2013N_Grzybowski,2005OS_Grier,2012OE_Tatiana} (FIG. 2a–d) or different liquid solvents (FIG. 2e–g), or, in the case of more complex crystals, by using bidisperse spheres \cite{2005N_Blaaderen,2011AFM_Vogel} (FIG. 2h–j), competing interactions \cite{2013PRL_Buzza} (for example, short-range attraction versus long-range repulsion) (FIG. 2k), templating \cite{2013PRL_Lowen3} or a combination thereof. After assembly, defects can be effectively annealed away by cycling tunable colloidal crystals near their melting points \cite{2012S_Han, 2010PRL_Han}, by shaking crystals in external fields \cite{2010PRL_Han,1999PRL_Maret,2004PRE_Wu} or with holographic optical tweezers \cite{2013PNAS_Chaikin}.

\subsection{Crystals composed of anisotropic particles}

Much of the interest in colloidal assembly comes from its potential applications for photonic crystals \cite{2011PUP_Robert}, and recent progress made in fabricating crystals composed of anisotropic particles is particularly valuable in this respect. Particle synthesis techniques have been used for the past three decades to produce micrometre-sized, non-spherical colloidal particles, such as rods, platelets, convex and concave polyhedrons, and spheres with bulges or dents \cite{1981ACR_Matijevic,1981JPS_Matsumoto, 2011COCIS_Lee,2011COCIS_Pine}; however, the phase-transition behaviour of such particles has rarely been studied. Recently, progress has been made in producing active particles \cite{2013RMP_Marchetti,2013JPCM_Sacanna}, anisotropic particles \cite{2013JPCM_Sacanna,2011COCIS_Pine,2013ACR_Granick} and particles with tunable interactions \cite{2006S_Dinsmore,2008N_Bechinger,2007SM_Yethiraj,2005S_Ayodh,2013NM_Mirkin,2014NM_Chaikin}. These experimental breakthroughs have greatly expanded our knowledge of anisotropic-particle assembly, and led to the computational prediction of diverse structures \cite{2012S_Glotzer,2013PRL_Dijkstra,2015ACP_Dijkstra}, although many of these have yet to be realized experimentally. Recent reviews of colloid fabrication are available in the literature \cite{2013JPCM_Sacanna, 2013ACR_Granick}; in this Review we focus on crystal assembly \cite{2015ACP_Dijkstra,2007NM_Glotzer,2011ACIE_Stein,2015ARPC_Granick}. 

Hard ellipsoids are a simple type of anisotropic particle, which can be stretched from spheres and can assemble into rotator crystals (that is, plastic crystals), liquid-crystal phases and crystals in 2D \cite{2014JCP_Odriozola} and 3D \cite{2013JCP_Odriozola}; however, experimental realization remains challenging because micrometre-sized ellipsoids tend to be trapped in glassy states \cite{2011PRL_Zheng}. By contrast, using an electric field, long-range repulsive rods can be tuned from a liquid to a rotator crystal \cite{2014NC_Blaaderen} (FIG. 3a) to a crystal without vitrification because of their abundant free space. Photolithography can produce numerous platelets with arbitrarily designed 2D shapes \cite{2007JPCC_Mason,2011PNAS_Mason,2014JPCM_Mason,2012NC_Mason}, which suffices for assembly in 2D but not in 3D. Monodispersed squares, triangles, pentagons and square crosses (FIG. 3b) were assembled into various 2D crystals by slightly tilting samples under gravity \cite{2011PNAS_Mason,2014JPCM_Mason,2012NC_Mason}. For example, square platelets form a hexagonal rotator phase and a rhombic crystal as density increases, but do not form any phase with four-fold symmetry \cite{2011PNAS_Mason}. Non-chiral triangles can form a crystal with left-handed and right-handed chiralities either mixed locally in experiments \cite{2011PNAS_Mason} or separated into two large domains in simulations \cite{2015SM_Dijkstra}. These platelets do not have tunable sizes or interactions; hence, it remains a challenge to drive a phase transition and study its kinetics. 

Polyhedra \cite{2011SM_Philipse,2014Angewandte_Blaaderen} have been fabricated and assembled into crystals (FIG. 3c) \cite{2012NM_Henzie}. The experimental study of colloidal crystals is more challenging in 3D than in 2D because it requires many more particles and because refractive-index matching between particles and the liquid solution is required for 3D confocal imaging. Hence, simulations take the leading role in studies of 3D assembly of anisotropic particles \cite{2015ACP_Dijkstra,2007NM_Glotzer}. Numerous types of crystals, plastic crystals, liquid crystals and amorphous phases were revealed in simulations for 145 types of hard polyhedrons by systematically varying their shapes \cite{2012S_Glotzer}. This demonstrates that a purely entropic effect can give rise to almost any structure observed in atomic systems, even for simple hard polyhedrons. 

\begin{figure}[!h]
\centering
\includegraphics[width=0.9\columnwidth]{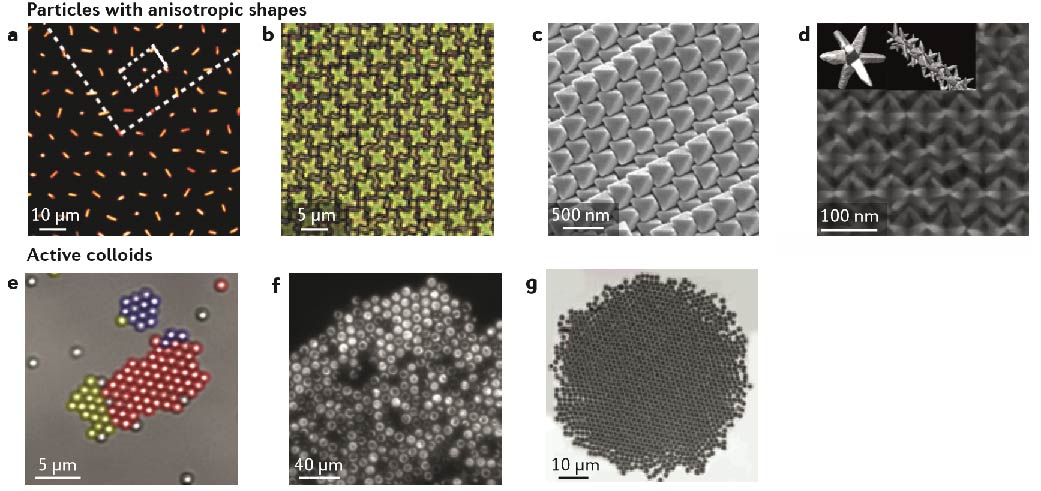}
\caption{\textbf{Colloidal crystals composed of anisotropic or active particles.}  (a) Rods with long-range repulsions form a rotator crystal with random orientations. (b) A 2D crystal assembled from square crosses. (c) Octahedra packing in 3D. (d) Hierarchical assembly of nano-octapods. (e) Self-propelled Janus spheres self-assemble into crystallites. (f) Spherical self-rotating Thiovulum majus bacteria self-assemble into a crystal with global random rotation. (g) Paramagnetic Janus spheres spin under a rotating magnetic field and form a crystallite. Panel a from REF.74, Nature Publishing Group. Panel b adapted with permission from REF.77, Institute of Physics. All rights reserved. Panel c from REF.185, Nature Publishing Group. Panel d from REF.82, Nature Publishing Group. Panel e adapted with permission from REF.111, American Association for the Advancement of Science. Panel f adapted with permission from REF.109, American Physical Society. Panel g reproduced with permission from REF.110, Royal Society of Chemistry.}
\label{fig:3}
\end{figure}

Compared with convex particles, concave particles can form more complex and even hierarchical structures \cite{2011NM_Manna}. For example, octapod-shaped nanoparticles interlock to form chains, which in turn stack to form a 3D super lattice with high porosity and rigidity \cite{2011NM_Manna} (FIG. 3d). Such hierarchical assembly greatly expands the possible structures. 

Besides shape anisotropy, interaction anisotropy can give rise to richer crystalline structures \cite{2015ACP_Dijkstra,2007NM_Glotzer,2015ARPC_Granick}, which in recent years have been made accessible by breakthroughs in the fabrication of patchy spheres \cite{2010Granick_AM} and DNA-coated particles \cite{2013NM_Mirkin}. A simple type of particle with anisotropic interactions is the Janus sphere, which comprises two hemispheres with different properties, for example, half latex and half metal \cite{2013CR_Granick}. In the past three decades, various coating methods for making Janus \cite{2013CR_Granick} and multipatch spheres \cite{2012N_Pine} have been developed. Patch-induced anisotropic interactions and different responses to external fields give rise to complex behaviour; however, novel lattices have only been experimentally achieved using triblock spheres with attractive hydrophobic patches. Surprisingly, these spheres formed a kagome lattice \cite{2011N_Granick} rather than a typical hexagonal lattice because of the rotational entropy of the spheres \cite{2013NM_Mao}. Simulations of various patchy particles have led to the discovery of new crystals \cite{2007JCP_Doye,2012JPCM_Sciortino,2012NC_Sciortino} and new phase diagrams \cite{2012JPCM_Sciortino,2012NC_Sciortino,2014SM_Kenneth} (FIG. 1f), and the extraction of some general features; for example, directional interactions can  promote hierarchical assembly \cite{2015ARPC_Jack,2013JCP_Dijkstra}.

Base-pair sequences in DNA can be programmed in biotechnology, allowing single-strand DNA to be used as glue with programmable base-pair binding sites. Although programmable anisotropic interactions are promising, micrometre-sized spheres have only been uniformly coated with DNA and thus have isotropic attractions. Colloidal spheres coated with complementary DNA sequences have temperature-sensitive attractions and can form crystallites \cite{2015S_Manoharan,2012NC_Crocker} (FIG. 2j). DNA-coated nanospheres can form large 3D crystals \cite{2014N_Mirkin}, but DNA-coated microspheres usually form aggregates or small crystallites \cite{2015S_Manoharan,2012NC_Crocker} because their attraction ranges relative to the repulsion cores are extremely short \cite{2015S_Manoharan,2012NC_Crocker}. In addition, directional binding lowers both energy and entropy, resulting in an abnormal free-energy minimum corresponding to small crystallites; further crystallite growth is prohibited by the higher free-energy barrier \cite{2015PNAS_Frenkel}. Hence, the global free-energy minimum of large crystals is kinetically inaccessible. Quantitative analysis \cite{2015PNAS_Frenkel,2014PRL_Frenkel} has led to the development of important protocols to fabricate large crystals, and experimental techniques are reaching maturity after two decades of \cite{2015S_Mirkin}. Consequently, DNA-coated spheres now have great potential to be assembled into new, designer structures (see REF.\cite{2015S_Mirkin} for a recent review on this subject). 

Modern simulation techniques can be used for efficiently exploring the large parameter space of anisotropic building blocks and their packing structures \cite{2012S_Glotzer, 2015ACP_Dijkstra,2007NM_Glotzer,2014SM_Kenneth,2011PRL_Dijkstra}, and, in turn, could reveal general rules governing their assembly and structure. In the past decade, simulations have played an important part in, and provided an indispensable understanding of, anisotropic particle assembly \cite{2015ACP_Dijkstra,2007NM_Glotzer}.

Besides colloids with anisotropic interactions, topological colloids in liquid crystals, such as rings \cite{2013N_Ivan} or spheres with topological interacting knots \cite{2011S_Muvsevivc}, represent a new class of interaction and assembly, and could form new crystalline structures \cite{2011S_Muvsevivc}.

\subsection{Crystals composed of active particles.}

`Active matter' contains self-driven units by converting stored or ambient energy into systematic movement, and includes animal groups, energy-consuming polymers and active colloids \cite{2013RMP_Marchetti}. Active particles are different from atoms: they can exhibit new phenomena and new states of matter. Active matter is a rapidly expanding area as a new platform to study non-equilibrium physics. In the past decade, colloids have been made active with translational or rotational self-propulsion by adding local chemical, temperature, electric or force gradients around individual particles. Examples include chemical-reaction-driven Janus particles, thermophoretic Janus particles, and the use of force gradients induced by surface tension or polymerization \cite{2015ARPC_Granick, 2010PCCP_Sen}. Self-propelled spheres can spontaneously form 2D triangular lattices (FIG. 3e) because a newly crystallized particle is buried by subsequently deposited particles before its rotation can align it in the direction needed to escape from the crystal \cite{2013PRL_Lowen}. This self-trapping mechanism is distinct from the crystallization of passive particles. In addition, self-rotating bacteria \cite{2015PRL_Libchaber} (FIG. 3f) and Janus paramagnetic spheres under a rotating magnetic field \cite{2015SM_Granick} (FIG. 3g) can assemble into a rotating crystal via hydrodynamic interactions. Current active-particle studies mainly focus on low-density 2D gases and small crystallites \cite{2013S_Chaikin}, whereas dense \cite{2015PRX_Cottin} and 3D active systems are desirable for phase-transition studies in the future.

\section{Phase Transitions}

The kinetics of a phase transition can affect the transition rate, the intermediate state and even the final structure; however, kinetic pathways are difficult to predict, owing to the complex high-dimensional free-energy landscape. Measurable kinetic pathways only exist near equilibrium, where a small number of collective coordinates are sufficient to describe the kinetic process. Quasi-equilibrium can be expected to hold if bond breaking and formation are much shorter than the nucleation timescale. First-order phase transitions typically follow the nucleation mechanism, which can be roughly divided into three stages: incubation, critical nucleus formation and growth of the post-critical nucleus. The small scales of the first two stages require measurement using optical microscopy (representative images of the different transitions are shown in FIGS. 4,5), whereas the third stage was often studied by light scattering in earlier colloid experiments \cite{1990CRC_Ackerson,1986PRL_Ackerson}. First-order transitions can be identified, for example, from the coexistence of two phases at equilibrium, a van der Waals loop in the free energy, hysteresis, an order-parameter jump or different asymptotic behaviours of susceptibility \cite{1987RPP_Binder}. Strictly speaking, phase transitions are defined at the thermodynamic limit. Most colloidal crystals in phase-transition studies contain millions of particles, thousands of which are in the field of view. Transitions within small colloidal crystallites can cast light on nanocrystal transformations, which are interesting in their own right.

\begin{tcolorbox}
\textbf{Gibbs free energy of nucleation:}\\
In classical nucleation theory, the Gibbs free energy of nucleation is expressed as:
\begin{equation} 
\Delta G = -V\rho \Delta \mu + A \gamma + E_{\textrm{strain}} - E_{\textrm{defect}}, 
\nonumber \end{equation}
where $V$: volume \\ 
$A$: area \\
$\gamma$: surface tension \\
$\rho$: number density of particles in the nucleus \\
$\Delta\mu$ ($>0$): chemical potential difference between the parent phase and the nucleus 
\\ $E_{\textrm{strain}}$: strain energy caused by volume change of the nucleus; zero for fluid parent phase and finite for solid parent phase \\
$E_{\textrm{defect}}$ ($>0$): energy of preexisting defects in volume $V$ of a crystalline parent phase
\end{tcolorbox}

CNT is most widely used to interpret experimental and simulation data \cite{2007JPCM_Sear}. In CNT (BOX 2), competing terms with opposite signs give rise to a barrier in $\Delta G$ (the Gibbs free energy of nucleation) that is a function of the nucleus size. Thus, small nuclei tend to shrink rather than grow, unless their size exceeds a critical value related to the barrier height of $\Delta G$ (red curve in FIG. 5a). The energy of pre-existing defects or surfaces ($E_defect$) reduces $\Delta G$. Thus, nucleation tends to start  heterogeneously from defects or surfaces.

CNT is based on certain assumptions. For example, it is assumed that the nucleus surface is rough and that growth takes place without the existence of preferential sites; that nuclei grow through monomer attachment rather than collective attachment of multiple particles; and that growth occurs about a single nucleus without coalescence of nuclei or precursors. These assumptions usually hold at weak super saturation; however, the deviations that occur at strong supersaturation11 are less clear. One research focus in nucleation is non-CNT behaviour, such as in the following three cases. First, the parent phase develops certain structures or dynamics before nuclei form. Second, the nucleus can evolve through intermediate metastable states with lower free-energy barriers rather than by directly transforming into the product phase \cite{2015ARPC_Jack,2014S_Samanta} (blue curve in FIG. 5a), which is called Ostwald's step rule. This rule, in principle, can be generally applied to any barrier-crossing process in a rugged landscape, such as protein folding and chemical reactions. In general, the intermediate metastable state is more likely to appear if particles have appropriate interaction ranges or possess different types of interactions that stabilize the different phases \cite{2015ARPC_Jack}. Third, instead of $\Delta G(n)$ for a nucleus with n particles in CNT, the nucleation may have additional collective coordinates, m (for example, crystallinity, number of particles in the intermediate phase \cite{2015PRL_Qi}, space curvature \cite{2014S_Manoharan}), which fall within multidimensional reaction-rate theory. The kinetics follows the steepest pathway through the saddle point of the free-energy surface $\Delta G(n,m)$ (FIG. 5b). Multiple paths can occur simultaneously \cite{2014NP_Tan} if their barrier heights are similar. More collective coordinates often yield a high-dimensional rugged landscape with many basins, similar to those shown in FIG. 5c. Note that basins may not be fully surrounded by barriers in higher dimensions; that is, they have narrow barrier-free channels connected to lower-energy states. Such morphology should be more prevalent for larger systems that have extremely high-dimensional landscapes. In addition to CNT, alternative theories can describe nucleation and provide different predictions \cite{2001PRE_Charles, 2013JCP_Miguel}. 

Tunable colloids enable us to drive a phase transition quasi-statically and cycle through the transition multiple times for better statistics. Temperature on a microscope stage usually cannot be tuned far from 300~K to change the thermal energy $k_BT$; hence, colloids are usually tuned by changing the particle interaction. Fabricating tunable colloids is one of the key breakthroughs of the past decade: examples are repulsive poly(N-isopropylacrylamide) (pNIPAM, NIPAM or NIPA) spheres with temperature-sensitive diameters \cite{2012S_Han, 2015NC_Han, 2005S_Ayodh,2010PRL_Han,2015NM_Han}; paramagnetic spheres \cite{1999PRL_Maret} or non-magnetic spheres in ferrofluids, for which dipolar interactions can be tuned by the magnetic field \cite{2015SM_Yellen}; temperature-sensitive DNA-binding \cite{2015S_Manoharan, 2012NC_Crocker,2014N_Mirkin,2011PNAS_Crocker,2014PNAS_Brenner}; fluctuation-induced \cite{2008N_Bechinger,2013NC_Schall,2015PRL_Dijkstra,2015NC_Sciortino} attractions; and depletion \cite{2006S_Dinsmore, 2015PNAS_Irvine, 2014NM_Chaikin} or surface absorption induced by temperature-sensitive polymers \cite{2014NM_Chaikin}. NIPA shells on latex cores create ellipsoidal, faceted or bowl-shaped particles with tunable size and shape \cite{2015NS_Schurtenberger}. Magnetic particles are ideal for assembling crystals in 2D but not in 3D, because the attractions along the magnetic field induce chains rather than crystals. Depletion-induced or fluctuation-induced attraction strength is often not finely tunable in the useful range of 0-1$k_BT$ for phase transitions. Adding a fraction of non-temperature-sensitive depletant can induce more finely tuned depletion attractions \cite{2015PNAS_Irvine}. 

Most experiments and simulations have been focused on colloidal phase transitions in either quasi-static processes or systems abruptly changed to a supersaturated state to observe the equilibration process. These results can be compared with CNT, which describes the evolution of a supersaturated system under fixed conditions. However, phase-transition kinetics has not been well studied while changing a thermodynamic variable, such as the heating or cooling rate, and the related dynamical theories are limited for first-order \cite{2014PRL_Eckstein} and second-order \cite{2014IJMPA_Campo} phase transitions.

\begin{figure}[!h]
\centering
\includegraphics[width=0.8\columnwidth]{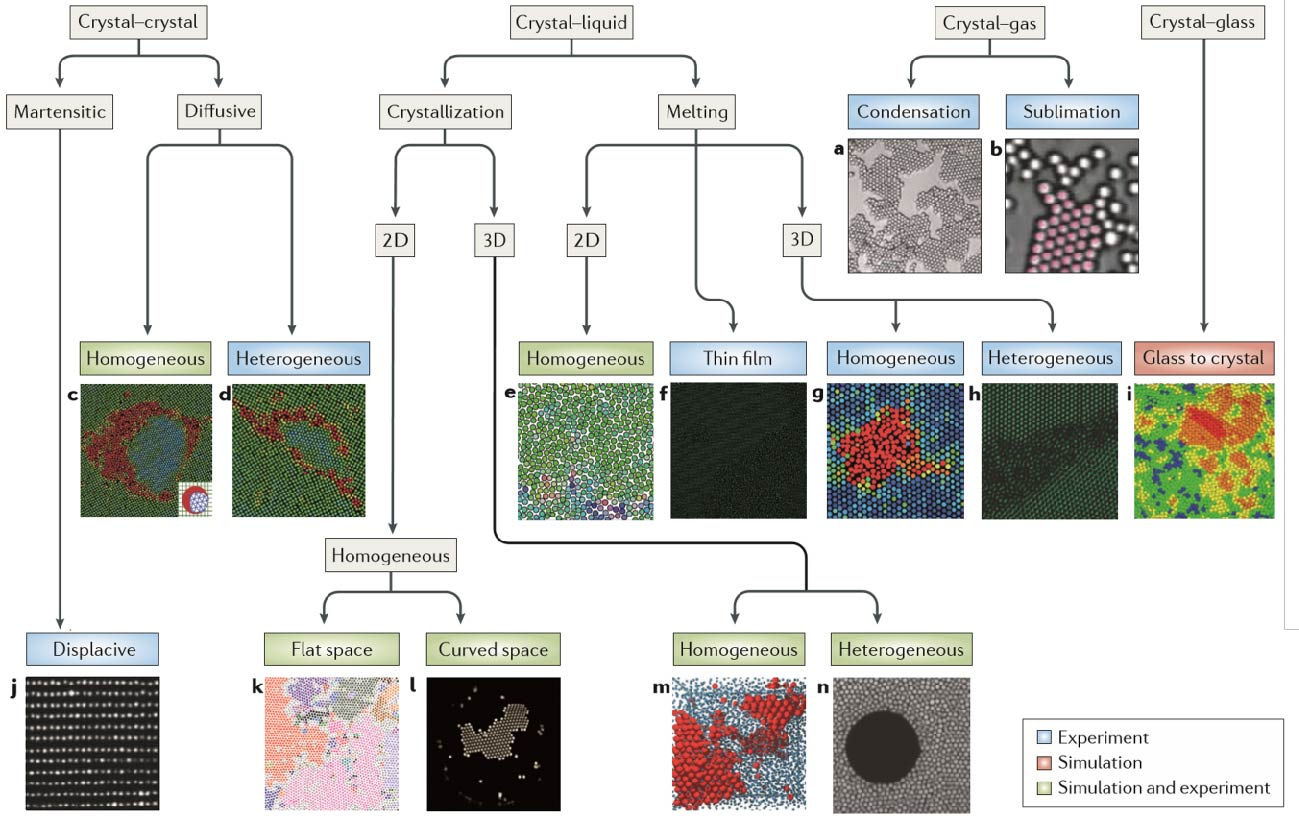}
\caption{\textbf{Phase-transition studies within colloidal crystals in the past two decades.} (a,b) Condensation and sublimation by tuning inter-particle attractions. (c,d) Transition from square to triangular lattice within a crystalline domain (panel c) and at a grain boundary (panel d). (e) 2D crystal melting. (f) Five-layer thin film melted from a dislocation and a grain boundary. (g) Melting in a defect-free region inside a 3D crystal. (h) Melting near a grain boundary inside a 3D crystal. (i) Crystallization of monodisperse hard-sphere glass. (j) Electric-field-induced fcc to bct transition. (k) Magnetic spheres crystallized to a polycrystal. (l) Crystallization on a droplet surface. (m) Crystallization inside 3D bulk supercooled liquid. (n) Crystallization from the surface of a large spherical seed. Image a from REF.64, Nature Publishing Group. Image b adapted with permission from REF.31, American Association for the Advancement of Science. Images c,d from REF.121, Nature Publishing Group. Image e adapted with permission from REF.154, American Physical Society. Image g adapted with permission from REF. 10, American Association for the Advancement of Science. Image h adapted with permission from REF.62, American Association for the Advancement of Science. Image i adapted with permission from REF.145, American Physical Society. Image j adapted with permission from REF.43, American Physical Society. Image k adapted with permission from REF.142, Proceedings of the National Academy of Science. Image l adapted with permission from REF.117, American Association for the Advancement of Science. Image m adapted with permission from REF.127, American Association for the Advancement of Science. Image n from REF.137, Nature Publishing Group.}
\label{fig:4}
\end{figure}

\subsection{Crystallization}

Crystallization is the most intensively studied phase transition in colloids (see reviews on this subject \cite{2009JPCM_Gasser,2014JPCM_Palberg} and references therein). Homogeneous crystallization (that is, crystallization without any preferential sites) of a 3D supercooled liquid can be observed deep inside the bulk, which avoids heterogeneous crystallization at walls. In a ground-breaking confocal microscopy experiment, the first in situ observation of 3D crystallization was achieved at the single-particle level by using hard-sphere colloids, thereby allowing measurements of the nucleus structure, size distribution, nucleation rate and surface tension to be taken \cite{2001S_Weitz}. The measured homogeneous nucleation rate was very different from the simulated values \cite{2001N_Frenkel}, which could be attributed to nucleation precursors \cite{2010PNAS_Tanaka} or other factors \cite{2007SM_Yethiraj,2014JPCM_Palberg}. Various nucleation precursors, such as the predicted bcc nucleus in transitions towards arbitrary product lattices \cite{1978PRL_Mctague}, intermediate-range ordered liquids \cite{2010PNAS_Tanaka}, dense liquid droplets \cite{2010PRL_Snook} and ordered clusters \cite{2014NP_Tan}, have been observed in different repulsive colloids. Crystallization in attractive colloids has been experimentally explored recently using DNA-coated spheres \cite{2015NC_Pine}. 
 
Heterogeneous crystallization in 3D has been studied on flat walls \cite{2003PRL_Frenkel,2004PRL_Dijkstra} and patterned substrates \cite{2012PRL_Granasy}, and for seed particles with different curvatures \cite{2004N_Frenkel,2015NC_Lowen}. The seed can be a large sphere \cite{2015NC_Lowen} or an array of optical traps \cite{2011SM_Blaaderen}. Small seeds suppress crystallization, whereas large seeds promote it \cite{2004N_Frenkel,2015NC_Lowen}. In particular, when the seed diameter is appropriate (several times that of the colloidal spheres), crystallites nucleate on the seed surface before detaching \cite{2004N_Frenkel,2015NC_Lowen} (FIG. 5d). 

2D freezing has been much less studied than 2D melting, although it can be easily observed by reversing the tuning parameters in 2D melting experiments. Note that melting and freezing share the same equilibrium phases, but their kinetics are different and beyond the scope of 2D melting theories \cite{1978PRL_Nelson,1979PRB_Young,1982PRL_Chui}. The kinetic Kibble–Zurek mechanism in cosmology \cite{2014IJMPA_Campo} successfully describes the scaling of the generation rates of topological defects, and the dynamical length scales at various quenching rates as a monolayer of paramagnetic colloidal spheres freezes \cite{2015PNAS_Keim}. This remarkable similarity between cosmology and soft-matter systems shows that certain features of the non-equilibrium kinetics of phase transitions are universal. 

Phase transitions in curved spaces are of basic interest, but they are difficult to study in atomic systems. They can, however, easily be observed in colloidal particles trapped at a fluid/fluid interface \cite{2014S_Manoharan,2010N_Chaikin}. Unlike crystallization in flat space, colloidal spheres on a liquid droplet surface nucleate into fractal-like crystalline patches (FIG. 5e) rather than a large crystal, because of curvature-induced strain \cite{2014S_Manoharan} (FIG. 5b). 

Deeply supercooled liquids have dynamic heterogeneity, in which the viscosity can differ by orders of magnitude in neighbouring mesoscopic regions. Such inhomogeneity would certainly affect crystallization, but has rarely been considered in nucleation theories. Simulations have revealed that crystallization and dynamic heterogeneity enhance each other in 3D hard-sphere supercooled liquids and glasses \cite{2009PRL_Pusey, 2011PRL_Cates}.

\subsection{Melting}

Although crystallization has been intensively studied, melting deserves separate investigation because the two processes differ in the following crucial aspects. First, melting and freezing points are usually different in colloids; for example, $\phi_\textrm{melt}=54.5\%$ and $\phi_\textrm{freeze}=49.4\%$ for hard-sphere crystals \cite{1986N_Megen} (FIG. 1a). At $49.4\%<\phi<54.5\%$, crystals with $\phi=54.5\%$ and liquids with $\phi=49.4\%$ coexist at equilibrium. Second, tunable colloids are required when the metastable parent phase is ordered (for example, during melting \cite{2012S_Han, 2005S_Ayodh}), but not for a disordered parent phase (for example, brute force can break crystals into supercooled liquids, which can equilibrate back via crystallization without tuning \cite{2001S_Weitz}). Third, defects are important in melting \cite{1988RMP_Katherine,2010PRL_Han,2005S_Ayodh}, but are generally insignificant in crystallization. Fourth, small nuclei can have various ordered structures in crystallization \cite{2014NP_Tan, 2009JPCM_Gasser,2014JPCM_Palberg,2010PNAS_Tanaka,1978PRL_Mctague}, but only disordered liquid exists in melting. Fifth, supercooled liquids are common in nature because crystallization can be easily suppressed below the freezing temperature, but superheated crystals \cite{2012S_Han} are rare because crystals will melt from free surfaces (that is, solid/vapour interfaces) once they reach the melting point. Sixth, the melting speed monotonically increases with the degree of superheating \cite{2015NC_Han}, whereas the crystallization rate is non-monotonic as the temperature decreases towards the vitrification point because of the increasing driving force of crystallization and the decreasing diffusivity of the particles \cite{2014JPCM_Palberg,2001N_Frenkel}. Seventh, nucleus expansion can induce strain in the parent lattice during melting, but crystallization induces no strain in the parent liquid phase. A larger liquid nucleus in melting causes more parent-lattice deformation and gives rise to a higher energy barrier. Hence, there is no kinetic pathway for melting when the driving force is weak under mild superheating \cite{2005PRB_Li} (black curve in FIG. 5a). This forbidden gap does not exist in crystallization. Last, after the liquid has fully crystallized, small polycrystalline grains coalesce into large ones \cite{2015PNAS_Keim}. Such a ripening process in crystallization does not exist in melting. 

Crystals tend to melt heterogeneously from pre-existing defects. The `defect strength' in terms of inducing melting decreases roughly in the following order: free surface (that is, solid/vapour interface) $>$ bulk triple junction $>$ high-angle grain boundary $>$ flat substrate $>$ low-angle grain boundary $>$ dislocation $>$ partial dislocation $>$ point defects (vacancy and interstitial). Stronger defects may also pre-empt melting at weaker defects. This expected trend has not, however, been well studied in colloidal systems. Heterogeneous melting at grain boundaries \cite{2005S_Ayodh} and hard walls \cite{2011JPCS_Han} has been experimentally observed in repulsive colloidal crystals. In nature, crystals are composed of attractive particles with free surfaces. Thus, they always melt from free surfaces because the pre-melted surface liquid serves as a huge post-critical nucleus. Surface pre-melting exists in almost all crystals \cite{2006RMP_Dash} and has been recently observed at the single-particle level using tunable attractive colloids \cite{2016N_Han}.

Homogeneous melting only occurs in defectfree crystalline regions if melting is avoided from surfaces and defects outside the region of interest. Homogeneous 3D melting was observed at the single-particle level for the first time by focusing a beam of light to superheat the interior of a NIPA colloidal crystal while the ambient crystal remained below the melting point\cite{2012S_Han}. During the incubation stage, particles swap positions with neighbours, and this triggers nucleation of melting. Such a nucleation precursor, rather than defects as is commonly assumed, keeps the lattice intact\cite{2012S_Han}. Particle swapping has also been observed in 2D crystals \cite{2014PNAS_Sprakel}. The nucleus growth kinetics agrees with CNT predictions under mild superheating, but deviates under strong superheating owing to non-CNT effects, such as nucleus shape fluctuations, nuclei coalescence and multimer attachment11. At the superheat limit where the superheated crystal crosses over from a metastable to an unstable state (blue dashed curve in FIG. 5a), the lattice becomes mechanically unstable and catastrophically breaks down\cite{2012S_Han} (FIG. 5g). This transition at the superheat limit is similar to the spinodal decomposition of crystallization at the supercool limit \cite{2009PR_Cavagna} and remains poorly understood. 

The predominant theory used to describe 2D melting is the Kosterlitz–Thouless–Halperin–Nelson– Young (KTHNY) theory, which predicts that 2D crystals melt via two continuous KT transitions through an intermediate hexatic phase \cite{1978PRL_Nelson,1979PRB_Young}. An alternative 2D melting scenario is the one-step first-order transition associated with the formation of grain boundaries \cite{1982PRL_Chui}. These theories do not exclude other melting scenarios. Many molecular monolayers exhibit one-step first- order melting \cite{1988RMP_Katherine}, but grain boundaries have not been reported in such systems in which the substrate effects are strong. By contrast, 2D colloidal monolayers confined between two flat walls \cite{2008PRE_Han} or at a fluid/fluid interface \cite{1999PRL_Maret} have negligible substrate effects, and defects can be directly observed. Melting of 2D colloidal crystals with tunable long-range \cite{1999PRL_Maret} or short-range \cite{2008PRE_Han} repulsions exhibits an intermediate hexatic phase, but the order of the two transitions cannot be well resolved experimentally. Simulations reveal that the melting behaviour can be correctly resolved only for systems containing more than 106 particles in 2D \cite{2006PRE_Mak,2015PRL_Krauth,2011PRL_Krauth}. Large-scale simulations show that disks with various repulsion ranges follow the KTHNY melting scenario except that the liquid–hexatic phase transition is weakly first-order for short-range repulsive disks \cite{2015PRL_Krauth,2011PRL_Krauth} (FIG. 1b).

Although 2D melting has been intensively studied in molecular and colloidal systems, the following three aspects are still poorly understood. First, the kinetics is beyond the scope of 2D melting theories \cite{1988RMP_Katherine} and has been poorly studied experimentally \cite{2015PNAS_Keim}. Second, 2D melting in attractive colloidal crystals has not been well studied \cite{2016N_Han}. Last, both KTHNY and grain-boundary- mediated melting theories are developed for infinitely large defect-free crystals without surfaces \cite{2006RMP_Dash}. Both 3D and 2D crystals with free surfaces are expected to premelt and melt from surfaces \cite{2006RMP_Dash}, although surface melting in 2D has only been indirectly measured in molecular monolayers by calorimetry \cite{1988PRB_Dash}.

Thin films are extremely valuable in technological applications, but their melting behaviour has not been thoroughly studied either theoretically or experimentally. A study of NIPA colloids confined between two walls \cite{2010PRL_Han,2011PRE_Han} showed that crystals comprising fewer than 5 layers melt catastrophically and homogeneously, 5- to 12-layer crystals melt heterogeneously from grain boundaries and dislocations (FIG. 5g), and $>$ 12-layer crystals only melt from grain boundaries, as is the case for 3D crystals.

\begin{figure}[!h]
\centering
\includegraphics[width=0.8\columnwidth]{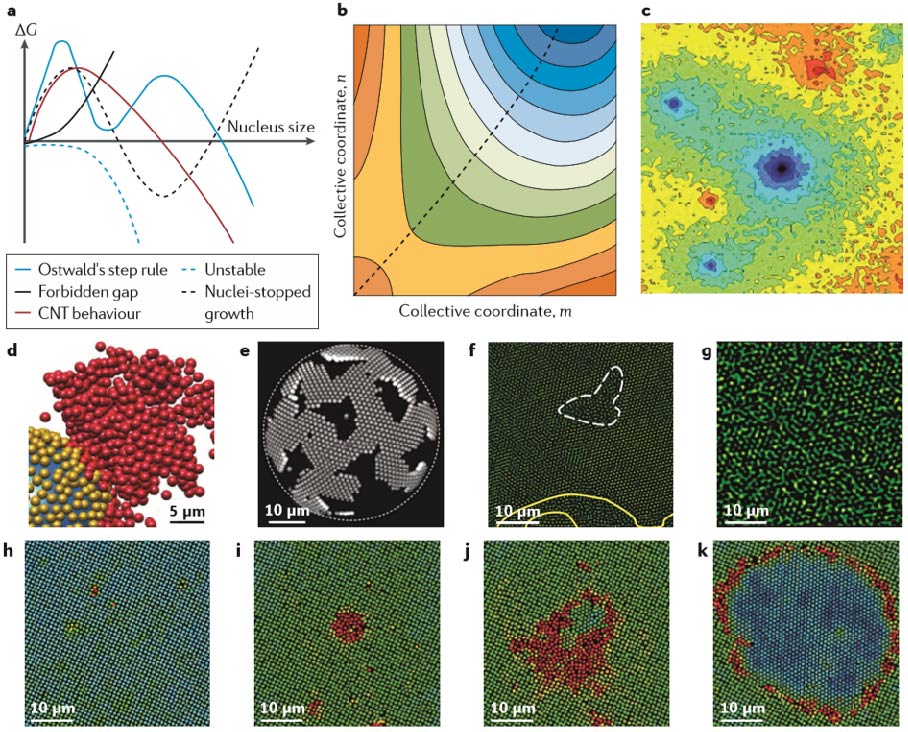}
\caption{\textbf{Phase-transition kinetics and free-energy barriers.} (a) Schematics of possible free-energy barriers. (b) Free-energy surface as a function of two collective coordinates (such as averaged width and length of crystalline arms in panel e). (c) Short-range attractive systems often have a rugged free-energy landscape with many local minima in basins depicted by closed contours. At low temperature, the system tends to be trapped in one of the basins (producing a disordered structure) rather than reaching the global minimum (crystal). (d) A crystallite (red particles) heterogeneously nucleates from the surface of a seed (large blue sphere) and then detaches from the seed. The image is reconstructed from particle positions. (e) Homogeneous crystallization in a curved 2D space (emulsion-droplet surface) results in a fractal-like crystal. (f) Heterogeneous melting from a grain boundary and a dislocation in a five-layer thin-film crystal. (g) Catastrophic melting at the superheat limit of a 3D crystal. (h-k) The transition from a square to a triangular lattice initially develops a metastable liquid nucleus, then a triangular lattice nucleates from within the liquid nucleus. CNT, classical nucleation theory; $\Delta G$, the change in Gibbs free energy. Panel d from REF.137, Nature Publishing Group. Panel e adapted with permission from REF.117, American Association for the Advancement of Science. Panel f adapted with permission from REF.157, American Physical Society. Panel g adapted with permission from REF.10, American Association for the Advancement of Science. Panels (h-k) from REF.121, Nature Publishing Group.}
\label{fig:5}
\end{figure}

\subsection{Solid-solid Transitions}

Solid–solid transitions between crystals prevail in the Earth's mantle and in metallurgy. They are more diverse, complex and less understood than crystal–liquid transitions. For example, water has 17 known crystalline structures but only one or two liquid phases. Solid–solid transitions are usually first-order phase transitions featuring an abrupt density change. Their kinetics typically follows either a diffusive nucleation or a martensitic transformation with concerted motion of particles. The microscopic kinetics and effect of dimensionality remain poorly understood. Most solid–solid transitions involve interparticle bond-breaking. Such reconstructive transitions are one of the least understood structural phase transitions because they are difficult to describe in theory (the difficulty lies in defining an order parameter because of the lack of a group–subgroup relation in the lattice symmetries) \cite{1996WS_Vladimir}; in simulations, owing to sluggish dynamics; or in experiments, owing to the lack of single-particle observations in the bulk. Hence, colloids are ideal for the study of solid–solid transitions. 

Solid–solid transitions have been far less studied than crystallization and melting in colloids, with most studies focused on structural properties either because the crystals were not tunable or the martensitic kinetics were too fast \cite{2004PRL_Blaaderen,2012NC_Crocker,2015SM_Yellen,1995JCP_Murray}. Solid–solid transitions in colloids have been driven by electric \cite{2004PRL_Blaaderen} and magnetic fields \cite{2015SM_Yellen} and temperature-sensitive DNA attractions \cite{2012NC_Crocker}. These systems show martensitic transformations with fast kinetics, promoted by global stretching of external fields or by small crystal size. Apart from these transitions with lattice symmetry changes, isostructural solid–solid transitions by abrupt lattice shrinking or dilation maintain the same lattice symmetry and often occur in metallic and multiferroic systems. Such transitions were discovered in simulations of attractive colloids \cite{1994PRL_Frenkel} and have been experimentally confirmed \cite{2016N_Han}. 

Diffusive solid–solid nucleation has recently been observed in superheated thin-film NIPA colloidal crystals under local optical heating. A new type of two-step nucleation process was discovered \cite{2015NM_Han}: 5-layer square lattice to liquid nucleus to 4-layer triangular lattice (FIG. 5h–k). The initial nucleus is liquid because small nuclei are dominated by surface tension, which is lower at the liquid/solid interface than at the solid/solid interface. This two-step nucleation occurs in defect-free regions, near dislocations and in grain boundaries, because the constant Edefect (BOX 2) does not change the shape of $\Delta G$. This provides a clear illustration of Ostwald's step rule (blue solid curve in FIG. 5a). The intermediate liquid has been observed in simulations of solid–solid transitions of hardsphere thin films \cite{2015PRL_Qi} and 2D ice \cite{2010PNAS_Zeng}, and may exist in other atomic systems because the liquid/solid surface tension is smaller than the solid/solid surface tension for most metals and alloys. Another diffusive transition was observed when the fcc lattice of soft NIPA spheres transforms into a bct lattice driven by an electric field; however, the reverse bct to fcc transition is martensitic \cite{2015PRX_Schurtenberger}. These kinetics have not previously been proposed in atomic solid–solid systems, and thus cast new light on solid–solid transition pathways.

\subsection{Other phase transitions}

Other phase transitions have been significantly less studied in colloids. Crystal sublimation \cite{2006S_Dinsmore,1996PRL_Grier} and reverse condensation \cite{2013NC_Schall,2009PRL_Dinsmore} occur only in attractive colloids. By weakening the attraction, 2D colloidal  crystals sublimate from surfaces. As the size decreases to ~30 particles, the whole crystallite disintegrates \cite{2006S_Dinsmore}.

All the phase transitions described in the previous sections were studied in isotropic colloids. Anisotropic particles exhibit rich phases, but studies of their phase transitions are limited; these include the crystallization of polyhedrons \cite{2014PRL_Escobedo}, self-propelled particles \cite{2012PRL_Lowen}, and solid–solid transitions between square and canted cubic crystallites \cite{2015PNAS_Irvine}. Active colloids show unique transitions \cite{2013PRL_Lowen2,2014PRL_Frey,2015NC_Tanaka} that do not exist in conventional passive thermal systems. For example, a transition from a resting to a travelling crystal in self-propelled spheres was predicted in simulations \cite{2013PRL_Lowen2}. In active colloids, 2D melting does not generate topological defects \cite{2014PRL_Frey} and involves much stronger dynamic heterogeneities than do thermal systems. It was found through simulations that adding hydrodynamic interactions induces a hexatic phase, rotator crystal or glass for active rotating disks \cite{2015NC_Tanaka}. Activecolloid experiments are crucial because it is challenging to take complex hydrodynamic interactions into account in simulations \cite{2015NC_Tanaka,2014PRL_Stark}. 

Non-equilibrium phase transitions that feature a breakdown of detailed balance \cite{2007Springer_Hinrichsen} are of great importance for both basic science and technology, yet are poorly understood. In colloids, non-equilibrium phase transitions have been briefly explored, including melting \cite{2009PNAS_Imhof} and crystallization under fluid flow \cite{2012PNAS_Cohen,2004N_Liu} or temperature gradients \cite{2011PRE_Schall}. Moreover, there have been a few studies of phase transitions in active colloids \cite{2013PRL_Lowen,2012PRL_Lowen,2014PRL_Frey}. Quantitative study of non-equilibrium phase transitions requires well-controlled conditions with a small number of collective coordinates.

\section{Perspective} 

Over the past two decades, colloidal crystals have provided remarkable insight into the microscopic kinetics of phase transitions. Studies have mainly been focused on bulk colloidal crystals composed of isotropic repulsive spheres, and on equilibrium and non-equilibrium behaviour at a fixed effective temperature. Thus, many aspects remain to be examined, for instance, the effects of surfaces, pre-existing defects and the small system size. In particular, phase transitions of surfaces and defects themselves, such as roughening transitions of free surfaces and grain boundaries, have not been explored. These aspects can be experimentally studied using currently available colloids and techniques. In metallurgy, controversies abound regarding defect dynamics, microstructure evolution and phase-transition kinetics because of the lack of single-particle resolution in the bulk. These can be studied in `colloidal metallurgy' because metallic atoms have relatively simple interactions. 

Many important processes remain to be explored in colloidal systems, including the kinetics during heating and cooling, transitions near critical points with strong fluctuations, transitions in anisotropic colloids and non-equilibrium transitions. Recent breakthroughs in fabricating anisotropic and active particles, tunable interactions and new assembly techniques provide vast possibilities for phase-transition studies. Currently, simulations take the leading role in the study of colloidal crystals composed of anisotropic or active particles, and more experimental studies are needed in the future. Researchers should try to derive general rules for how a particle's shape, interactions and activity affect crystal assembly and phase transitions. These general rules would help to achieve the goal of reverse engineering, that is, creating building blocks that will self-assemble into a prescribed structure \cite{2015S_Mirkin}, which is of fundamental importance for materials science. 

Compared with the commonly used bottom-up assembly, which inevitably results in defects in crystals, more powerful top-down fabrication can directly `print' crystals without defects or with defects in any designed pattern. Micro-platelets can be printed in ordered 2D arrays by lithography \cite{2007JPCC_Mason}, but washing platelets away from the substrate disrupts the crystalline structure. Non-destructive washing would be particularly desirable in the future. In addition, current 3D printing can be performed in a liquid environment at 50-$\mu$m resolution \cite{2015S_DeSimone}. The rapidly evolving 3D printing and etching technologies may give rise to a revolutionary top-down approach to fabricating colloids and colloidal crystals. New particle-tracking algorithms \cite{2014NMethod_Gonzalo}, super-resolution microscopy \cite{2015ACIE_Moerner}, liquid-environment transmission-electron microscopy \cite{2015S_Alivisatos} and, in particular, fast light-sheet microscopy \cite{2015NMethod_Editorial,2014S_Betzig} could also greatly benefit phase-transition studies if implemented in colloid experiments.

This work was supported by RGC-GRF16301514 and ANR/RGC-A-HKUST616/14 grants.

\newpage

\end{document}